\documentclass[twocolumn,english,superscriptaddress,secnumarabic,amssymb,amsmath,nobibnotes,aps,prd,showkeys,showpacs]{revtex4}
\usepackage[latin9]{inputenc}
\usepackage{amsmath}
\usepackage{amssymb}
\usepackage{graphicx}
\usepackage{esint}

\makeatletter
\@ifundefined{textcolor}{}
{%
 \definecolor{BLACK}{gray}{0}
 \definecolor{WHITE}{gray}{1}
 \definecolor{RED}{rgb}{1,0,0}
 \definecolor{GREEN}{rgb}{0,1,0}
 \definecolor{BLUE}{rgb}{0,0,1}
 \definecolor{CYAN}{cmyk}{1,0,0,0}
 \definecolor{MAGENTA}{cmyk}{0,1,0,0}
 \definecolor{YELLOW}{cmyk}{0,0,1,0}
}

\usepackage{epsfig}\usepackage{amstext}\usepackage{bm}
\usepackage{booktabs}\usepackage{color}
 
\let\baraccent=\= 
\renewcommand{\=}[1]{\stackrel{#1}{=}} 

\usepackage{babel}

\makeatother

\usepackage{babel}
\begin{document}

\title{Isofrequency pairing of spinning particles in Schwarzschild-de Sitter
spacetime}

\author{Daniela Kunst}

\email{daniela.kunst@zarm.uni-bremen.de}

\affiliation{ZARM, University of Bremen, Am Fallturm, 28359 Bremen, Germany}

\author{Volker Perlick}

\email{volker.perlick@zarm.uni-bremen.de}

\affiliation{ZARM, University of Bremen, Am Fallturm, 28359 Bremen, Germany}

\author{Claus L{ä}mmerzahl}

\email{claus.laemmerzahl@zarm.uni-bremen.de}

\affiliation{ZARM, University of Bremen, Am Fallturm, 28359 Bremen, Germany}

\affiliation{Institute for Physics, University Oldenburg, 26111 Oldenburg, Germany}
\begin{abstract}
It has been established in Schwarzschild spacetime (and more generally
in Kerr spacetime) that pairs of geometrically different timelike
geodesics with the same radial and azimuthal frequencies exist in
the strong field regime. The occurrence of this socalled isofrequency
pairing is of relevance in view of gravitational-wave observations.
In this paper we generalize the results on isofrequency pairing in
two directions. Firstly, we allow for a (positive) cosmological constant,
i.e., we replace the Schwarzschild spacetime with the Schwarzschild-de
Sitter spacetime. Secondly, we consider not only spinless test-particles
(i.e., timelike geodesics) but also test-particles with spin. In the
latter case we restrict to the case that the motion is in the equatorial
plane with the spin perpendicular to this plane. We find that the
cosmological constant as well as the spin have distinct impacts on
the description of bound motion in the frequency domain. 
\end{abstract}
\maketitle

\section{Introduction}

\label{sec:Intro} 

The occurrence of \textit{isofrequency pairs} in the strong field
regime of the Schwarschild spacetime was noticed only recently by
Barack and Sago \cite{Barack11}. It says that bound orbits, for which
both the radial and the azimuthal motion are periodic, are not uniquely
characterized by their frequencies. More precisely, there exist geometrically
distinct orbits in the strong field regime that possess the same frequency
pairs.

At first, this degeneracy feature may not be much of a surprise. After
all, it is known from Newtonian Mechanics that the frequencies of
the Kepler ellipses are all degenerate, i.e. the radial and azimuthal
frequencies have the same value. This is the reason why the orbits
in Newtonian physics are closed.

When general relativistic effects are considered, though, one major
difference to Newtonian physics is the periastron shift of bound orbits
which are no longer closed. This manifests itself in the non-degeneracy
of the frequencies. In the Schwarzschild spacetime we have two independent
orbital frequencies, for the radial and for the azimuthal motion.
It was long thought that these two frequencies provide another unique
parametrization of the orbits, as an alternative to the ones already
known, such as the energy and the angular momentum or the periastron
and the apastron. However, Barack and Sago \cite{Barack11} showed
that in the strong field of Schwarzschild spacetime, i.e. in the highly
relativistic regime, there exist pairs of timelike geodesics which
are described by the same frequencies. In a follow-up study, Warburton,
Barack and Sago \cite{Warburton13} generalized the isofrequency pairing
to the Kerr geometry. In contrast to the Schwarzschild case, timelike
geodesics in the Kerr spacetime have three degrees of freedom and
in their study Warburton et al. found even triperiodic partners. As
outlined in Refs. \cite{Barack11,Warburton13}, the occurrence of
isofrequency pairing is of relevance in view of gravitational wave
analysis because it implies that, for the case of an Extreme Mass
Ratio Inspiral (EMRI), from the observation of the fundamental frequencies
one cannot uniquely determine the shape of the orbit. Shortly after
Refs. \cite{Barack11,Warburton13} had appeared, Shaymatov, Atamurotov
and Ahmedov \cite{Shaymatov14} investigated the influence of a magnetic
field on the orbital frequencies of a charged particle moving in Schwarzschild
spacetime and found that the region where isofrequency pairing occurs
shrinks for high values of the magnetic field. 

In this paper we generalize the results on Schwarzschild isofrequency
pairing in two different directions. Firstly, we consider the Schwarzschild-de
Sitter spacetime, which is the unique spherically symmetric and static
solution to Einstein's vacuum field equations with a positive cosmological
constant. Secondly, we consider not only spinless but also spinning
test paricles. We restrict to the case that the particle moves in
the equatorial plane, with the spin perpendicular to this plane. Properties
of the Schwarzschild-de Sitter spacetime have been discussed, e.g.,
in \cite{Stuchlik99-2,Mortazavimanesh09,Obukhov11}, and the motion
of non-spinning test-particles in this spacetime has been investigated,
e.g., in \cite{Stuchlik99,Stuchlik07,Hackmann08}. For spinning particles
we have to consider the equations of Mathisson-Papapetrou-Dixon \cite{Mathisson37,Papapetrou51,Dixon70}.
As this set of equations is not closed in the sense that there are
less equations than variables, an additional condition is needed,
usually referred to as a spin supplementary condition (SSC). Choosing
an SSC is associated with fixing a reference frame the motion is observed
in. The choice of the SSC depends on the question one wants to investigate
\cite{Semerak99,Kyrian07,Corinaldesi51,Georgios,NewtonWigner49,Tulczyjew59,Pirani56,Moeller49,Costa12}.
We use the Tulzcyjew condition (T SSC) \cite{Tulczyjew59} which corresponds
to the zero 3-momentum frame. The dynamics of spinning test-particles
has been worked out in Lagrangian and Hamiltonian formalisms of which
the latter appears to be more complicated \cite{Bailey75,Steinhoff11,Steinhoff08}.
For the motion of spinning test-particles in the Schwarzschild or
Kerr spacetime we refer, e.g., to \cite{Hackmann14,Plyatsko12,Plyatsko13}
and for the case that also the quadrupole moment of the test-particle
is taken into account to \cite{Steinhoff12}. Several papers have
been devoted to the question of whether the motion is chaotic, which
also has a great impact on the analysis of possible gravitational
wave signals \cite{Hartl03a,Hartl03b,Suzuki97,Verhaaren10}.

The paper is organized as follows. In the second section we recall
the equations of motion for non-spinning and spinning test-particles
in Schwarzschild-de Sitter spacetime. In particular we specify the
region of bound motion. After that we analyze the frequencies of bound
motion and find the domain where isofrequent orbits exist in section
three. Lastly, we summarize our results and give an outlook for future
work and applications.

\section{Motion in Schwarzschild de Sitter Spacetime}

\label{sec:SdS}

\subsection{Non-spinning particles }

Schwarzschild-de Sitter spacetime is the unique spherically symmetric
vacuum solution to Einstein's field equations including a positive
cosmological constant, $\Lambda>0$, see e.g. Rindler \cite{Rindler01}.
For a negative cosmological constant it is called Schwarzschild-anti-de
Sitter spacetime, but here we are interested only in the case $\Lambda>0$.
The metric reads \cite{Rindler01}

\begin{equation}
\small ds^{2}=-f\left(r\right)dt^{2}+f\left(r\right)^{-1}dr^{2}+r^{2}\left(d\theta^{2}+\sin\left(\theta\right)^{2}d\phi^{2}\right)\label{eq:metric}
\end{equation}
in spherical coordinates with 
\begin{equation}
f\left(r\right)=1-\frac{2M}{r}-\frac{\Lambda}{3}r^{2}\,.\label{eq:f}
\end{equation}
Here $M$ corresponds to the mass of the gravitating body and the
units are chosen such that $c=1$ and $G=1$. We consider only the
region $r>0$. For $0<\Lambda<(3M)^{-2}$ , the function $f(r)$ has
two zeros, indicating two horizons, at radii $r_{H1}$ and $r_{H2}$
with $2M<r_{H1}<3M<r_{H2}$; the metric is static in the region $r_{H1}<r<r_{H2}$
where $f(r)>0$. For $\Lambda>(3M)^{-2}$ there is no horizon and
no static region.

The dynamics of massive non-spinning test-particles is determined
by the geodesic equation for timelike curves. Since the spacetime
is spherically symmetric the motion takes place in a plane, which
we choose to be the equatorial plane $\theta=\frac{\pi}{2}$. Then
the equations of motion, parametrized by proper time $\tau$, for
a test-particle of mass $m$ read 
\begin{eqnarray}
m^{2}\Big(\frac{dt}{d\tau}\Big)^{2} & = & \frac{H^{2}}{f(r)^{2}}\,,\label{eq:teom}\\
m^{2}\Big(\frac{dr}{d\tau}\Big)^{2} & = & H^{2}-f(r)\left(\frac{J_{z}^{2}}{r^{2}}+m^{2}\right)\,,\label{eq:reom}\\
m^{2}\Big(\frac{d\phi}{d\tau}\Big)^{2} & = & \frac{J_{z}^{2}}{r^{4}}\,,\label{eq:phieom}
\end{eqnarray}
with $H$ and $J_{z}$ being two constants of motion to be interpreted
as the energy and the angular momentum, respectively. The particle's
four-momentum is $p^{\mu}=m\frac{dx^{\mu}}{d\tau}$ and satisfies
the mass-shell condition $m^{2}=-p_{\mu}p^{\mu}$. From hereon we
rescale $H\mapsto Hm$ and $J_{z}\mapsto J_{z}m$ which is tantamount
to setting $m=1$ in \eqref{eq:teom}, \eqref{eq:reom} and \eqref{eq:phieom}.
Then $H$ is dimensionless while $J_{z}$ has the dimension of a length.

In the following we are interested only in bound motion, i.e., in
orbits that have two turning points, an apastron $r_{a}$ and a periastron
$r_{p}$, where $dr/d\tau=0$. From (\ref{eq:reom}) it is obvious
that turning points can exist only at radial coordinates where $f(r)>0$.
This implies that bound orbits are confined to the region between
the two horizons. In particular they do not exist if $\Lambda>\left(3M\right)^{-2}$.

We rewrite (\ref{eq:reom}) as 
\begin{equation}
\Big(\frac{dr}{d\tau}\Big)^{2}=\frac{\Lambda}{3r^{3}}\, P_{5}\left(r\right)\,,\label{eq:reomP5}
\end{equation}
where 
\begin{eqnarray}
P_{5}\left(r\right) & =r^{5}-\left(\big(1-H^{2}\big)\dfrac{3}{\Lambda}-J_{z}^{2}\right)r^{3}\nonumber \\
 & +\dfrac{6M}{\Lambda}r^{2}-\dfrac{3}{\Lambda}J_{z}^{2}r+\dfrac{6M}{\Lambda}J_{z}^{2}\,.\label{eq:P5}
\end{eqnarray}
As we assume $\Lambda>0$, the region where $P_{5}(r)<0$ is forbidden
by (\ref{eq:reomP5}). The number of zeros of $P_{5}(r)$ determines
the types of motion possible for the corresponding values of $H$
and $J_{z}$. For positive $\Lambda$ there can be at most four positive
real zeros as can be derived with the rule of signs by Descartes.
Bound orbits are allowed only when the polynomial has precisely four
positive zeros. For other values of $H$ and $J_{z}$ there will be
escape or terminating orbits or no motion at all.

From now on we rescale $r\mapsto rM$, $J_{z}\mapsto J_{z}M$ and
$\Lambda\mapsto\Lambda M^{-2}$ so that these quantities are dimensionless.
(Recall that $H$ already was dimensionless.) This is tantamount to
setting $M=1$ in \eqref{eq:P5}. Since we must have four positive
real zeros in order to consider bound motion we can rewrite equation
\eqref{eq:P5} as 
\begin{equation}
P_{5}(r)=(r-r_{0})(r-r_{1})(r-r_{p})(r-r_{a})(r-r_{2})\label{eq:Vr}
\end{equation}
with $r_{0}<0<r_{1}<r_{p}<r_{a}<r_{2}$. Bound motion exists only
between $r_{p}$ and $r_{a}$. Also, the motion is fully described
by the parameters $r_{a}$ and $r_{p}$, because $H$ and $J_{z}$
can be expressed in terms of $r_{a}$ and $r_{p}$, 
\begin{eqnarray}
H^{2} & = & \frac{\left(\frac{\Lambda}{3}r_{a}^{3}-r_{a}+3\right)\left(\frac{\Lambda}{3}r_{p}^{3}-r_{p}+3\right)(r_{a}+r_{p})}{r_{a}^{2}(r_{p}-2)+r_{p}^{2}\left(r_{a}-2\right)-2r_{a}r_{p}}\,,\label{eq:Hrarp}\\
J_{z}^{2} & = & \frac{r_{a}^{2}r_{p}^{2}(3-\frac{\Lambda}{3}r_{a}r_{p}(r_{a}+r_{p}))}{r_{a}^{2}(r_{p}-2)+r_{p}^{2}\left(r_{a}-2\right)-2r_{a}r_{p}}\,.\label{eq:Jrarp}
\end{eqnarray}
Therewith we can find all the zeros of \eqref{eq:P5} expressed in
terms of $r_{p}$ and $r_{a}$. Comparing the coefficients of the
two equations \eqref{eq:P5} and \eqref{eq:Vr} yields expressions
for $r_{0}$ and $r_{2}$ dependent on $(r_{1},r_{a},r_{p})$ and
the equation 
\begin{eqnarray}
P_{3}(r_{1}):=\Lambda r_{a}^{2}r_{p}^{2}r_{1}^{3}+\Lambda r_{a}^{2}r_{p}^{2}(r_{a}+r_{p})r_{1}^{2}\nonumber \\
+J_{z}^{2}(6r_{p}-3r_{a}(r_{p}-2))r_{1}+6J_{z}^{2}r_{a}r_{p} & \stackrel{!}{=} & 0
\end{eqnarray}
so that we only have to solve a cubic equation for $r_{1}$.

In the $(H,J_{z})-$plane, the boundaries of bound motion are determinded
by the merging of two zeros or by the condition that the orbit becomes
unbound. Physically, a merging $r_{a}=r_{p}$ corresponds to a stable
circular orbit. A merging $r_{1}=r_{p}\neq r_{a}\neq r_{2}$ corresponds
to an unstable circular orbit at $r_{p}$ and a homoclinic orbit from
$r_{p}$ to $r_{a}$ and back to $r_{p}$, while a merging $r_{1}\neq r_{p}\neq r_{a}=r_{2}$
corresponds to an unstable circular orbit at $r_{a}$ and a homoclinic
orbit from $r_{a}$ to $r_{p}$ and back to $r_{a}$. In the case
that $r_{1}=r_{p}\neq r_{a}=r_{2}$ we have unstable circular orbits
at $r_{p}$ and at $r_{a}$ and heteroclinic orbits from $r_{a}$
to $r_{p}$ and from $r_{p}$ to $r_{a}$.

Another parametrization of bound orbits is given by the semi-latus
rectum $p$ and the eccentricity $e$. The relations to the periastron
$r_{p}$ and the apastron $r_{a}$ are 
\begin{equation}
r_{a}=\frac{p}{1-e}\,,\qquad r_{p}=\frac{p}{1+e}\,.\label{eq:rarppe}
\end{equation}
We choose this parametrization since then the analysis of the properties
of a test particle's motion is more convenient.

\begin{figure}[tph]
\center{\includegraphics[width=7cm]{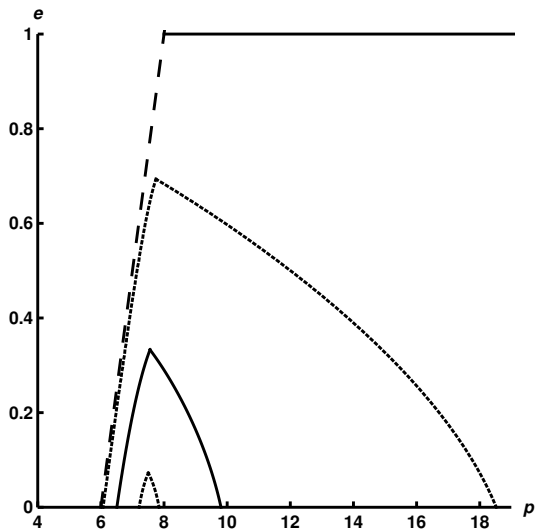}}
\caption{The figure presents the regions of bound motion in the $(p,e)$ -
plane for different values of $\Lambda$. The dashed line is the separatrix
in the Schwarzschild case, $\Lambda=0$, given by the equation $p=6+2e$.
The remaining lines are the separatrices for a cosmological constant
of $\Lambda=0.0001$ (upper dotted line), $\Lambda=0.0005$ (solid
line), and $\Lambda=0.0006$ (lower dotted line). }

\label{Fig:Seppe} 
\end{figure}

\vspace{0cm}

In Fig.~\ref{Fig:Seppe} the region of bound motion is shown in a
$(p,e)$-diagram for both Schwarzschild and Schwarzschild-de Sitter
spacetimes. We first consider the Schwarzschild case for which the
general features of the domain of bound orbits have been discussed
already by Cutler, Kennefick and Poisson \cite{Cutler94}, see also
\cite{Glampedakis02} and \cite{Barack11,Warburton13}. In this case
the domain of bound orbits, corresponding to the region that lies
to the right of the dashed line, is infinitely large. This dashed
line, which is given by the equation $p=6+2e$, is often called the
\emph{separatrix} because it separates the region of bound orbits
from the region of unbound orbits. On the separatrix, each point corresponds
to a homoclinic orbit from a radius $r_{1}=r_{p}$ to a radius $r_{a}>r_{p}$
and back to $r_{1}=r_{p}$. Such a homoclinic orbit has the same constants
of motion $H$ and $J_{z}$ as the unstable circular orbit at $r_{1}=r_{p}$
which it asymptotically approaches. 
Bound orbits near the separatrix are periodically ``zooming out''
to the apastron with many ``whirls'' near the periastron in between,
see Levin, O'Reilly and Copeland \cite{Levin00}; such \emph{zoom-whirl}
orbits have also been discussed in \cite{Glampedakis02,Levin09,Shaughnessy03}.
The upper boundary curve of the region of bound orbits corresponds
to unbound (parabolic-type) orbits with $e=1$, while the lower boundary
corresponds to stable circular orbits with $e=0$. The lower left-hand
corner of the region of bound orbits corresponds to the innermost
stable circular orbit (ISCO) at $r=6$.

By contrast, in the Schwarzschild-de Sitter spacetime the region of
bound orbits in the $(p,e)-$plane is finite. The shape is triangle-like
with its tip at an eccentricity $e_{\mathrm{max}}<1$. This demonstrates
that, in this picture, the transition from bound orbits to unbound
orbits ($e>1$) is not continuous. In analogy to the Schwarzschild
case, the two sides of the triangle are called the \emph{separatrices}.
They correspond to homoclinic orbits with $r_{1}=r_{p}$ and $r_{a}=r_{2}$,
respectively. Each homoclinic orbit has the same values for energy
and angular momentum as the unstable circular orbit that it approaches
asymptotically. 
We have already mentioned that in the Schwarzschild spacetime bound
orbits near the separatrix show a zoom-whirl behavior, with the ``whirling''
taking place in the strong-field regime. In the Schwarzschild-de Sitter
spacetime with a small positive cosmological constant, the second
separatrix gives rise to zoom-whirl orbits that ``whirl'' near an
apastron far away from the center and periodically ``zoom in'' to
a periastron. The two separatrices intersect at the tip of the triangle
where we have simultaneously $r_{1}=r_{p}$ and $r_{a}=r_{2}$. This
gives rise to a heteroclinic orbit, i.e., to an orbit that connects
two different unstable circular orbits. Such an orbit can be described
by a unique pair of values for $(p,e)$. The lower boundary curve
of the region of bound orbits corresponds to stable circular orbits.
The two intersection points of the separatrices with the horizontal
axis correspond to the innermost stable circular orbit (ISCO) and
the outermost stable circular orbit (OSCO). While the ISCO is also
present in the Schwarzschild spacetime and merely gets shifted away
from the center with increasing $\Lambda$, the OSCO is only existent
in the Schwarzschild-de Sitter spacetime. 
For $\Lambda\rightarrow0$ the OSCO approaches infinity. For $\Lambda\rightarrow\Lambda_{\mathrm{crit}}=4/5625$
the ISCO and the OSCO merge into one circular orbit, cf. \cite{Stuchlik99}.
Put into mathematical language, this happens if all four positive
zeros of \eqref{eq:P5} coincide. For $\Lambda>\Lambda_{\mathrm{crit}}$
bound orbits do not exist. Let us remark here that $\Lambda_{\text{crit}}$
is much larger than the physically expected value of the cosmological
constant. Observations show evidence for a $\Lambda\approx10^{-52}\mathrm{m}^{-2}$
\cite{Peebles03}. Even for a supermassive black hole with $M\approx10^{10}\mathrm{km}$
this corresponds, in our geometrized units, to $\Lambda\approx10^{-25}$,
i.e., to a value that is much smaller than $\Lambda_{\mathrm{crit}}$.

\subsection{Spinning particles in the equatorial plane}

Now we turn on the spin and investigate the motion of a spinning test-particle
which is no longer geodesic. The corresponding equations of motion
are the well-known Mathisson-Papapetrou-Dixon equations \cite{Mathisson37,Papapetrou51,Dixon70},
\begin{align}
\dot{p}^{\mu}= & -\frac{1}{2}R_{\,\nu\kappa\lambda}^{\mu}u^{\nu}S^{\kappa\lambda}\,,\label{eq:pMPD}\\
\dot{S}^{\mu\nu}= & \,\, pu^{\nu}-p^{\nu}u^{\mu}\,.\label{eq:SMPD}
\end{align}
Here $R_{\,\,\,\,\,\nu\kappa\lambda}^{\mu}$ is the curvature tensor
of the underlying spacetime, the dot denotes differentiation with
respect to proper time $\tau$ along the worldline of the particle,
$p^{\mu}$ is the four-momentum and $u^{\mu}$ is the four-velocity
of the particle, and $S^{\mu\nu}$ is the antisymmetric spin tensor.
As already mentioned in the introduction, the set \eqref{eq:pMPD}-\eqref{eq:SMPD}
is underdetermined; a so-called \emph{supplementary spin condition}
(SSC) must be chosen in order to get a well-posed initial-value problem.
In this work the Tulzcyjew condition \cite{Tulczyjew59} is chosen,
\begin{equation}
S^{\mu\nu}p_{\nu}=0\,.\label{eq:TSSC}
\end{equation}
Then the relation between $p_{\mu}$ and $u^{\nu}$ can be given explicitly:
Using the conserved quantity $m^{2}=-p_{\mu}p^{\mu}$, introducing
the normalized momentum vector $\hat{p}^{\mu}=\frac{p^{\mu}}{m}$
and renormalizing the four-velocity such that $\hat{p}^{\mu}\hat{u}{}_{\mu}=-1$
allows to derive the relation \cite{Saijo98} 
\begin{equation}
\hat{u}{}^{\mu}-\hat{p}^{\mu}=\frac{S^{\mu\nu}R_{\nu\gamma\sigma\lambda}\hat{p}^{\gamma}S^{\sigma\lambda}}{2\left(m^{2}+\frac{1}{4}R_{\alpha\beta\delta\xi}S^{\alpha\beta}S^{\delta\xi}\right)}\,.\label{eq:pu}
\end{equation}
Therewith, it is now possible to derive the equations of motion for
a spinning test-particle in Schwarzschild-de Sitter spacetime. However,
here we will restrict ourselves to the special case of a particle
moving in the equatorial plane, with the spin vector perpendicular
to this plane. Then we can characterize the spin by the scalar constant
of motion $S$, defined by 
\begin{equation}
S^{2}=S^{\mu}S_{\mu}\,,\qquad S^{\mu}=\dfrac{1}{2}\varepsilon^{\mu\nu\sigma\tau}\hat{p}{}_{\nu}S_{\sigma\tau}\label{eq:defs}
\end{equation}
with the totally antisymmetric Levi-Civita tensor field, and the property
that $S$ is positive if the spin is parallel to the orbital angular
momentum and negative if it is anti-parallel. As in the spinless case,
we have a conserved energy $H$ and a conserved angular momentum $J_{z}$
which we rescale according to $H\mapsto Hm$ and $J_{z}\mapsto J_{z}m$.
In addition, we now also rescale the spin, $S\mapsto sm$. The resulting
equations of motion can be taken from \cite{Saijo98}, after appropriate
adaptation, 
\begin{eqnarray}
\Big(\frac{dt}{d\tau}\Big) & = & \frac{H+\frac{f'(r)}{2r}sJ_{z}}{\Pi_{s}(r)\Sigma_{s}(r)f(r)}\,,\label{eq:teomS}\\
\Big(\frac{dr}{d\tau}\Big) & =\pm & \frac{\sqrt{R_{s}(r)}}{\Pi_{s}(r)\Sigma_{s}(r)}\,,\label{eq:reomS}\\
\Big(\frac{d\phi}{d\tau}\Big) & = & \frac{\big(J_{z}-Hs\big)\left(1-\frac{f''(r)}{2}s^{2}\right)}{\Pi_{s}(r)\Sigma_{s}(r)^{2}r^{2}}\,.\label{eq:phieomS}
\end{eqnarray}
Here $f(r)$ is defined by (\ref{eq:f}), the prime denotes derivative
with respect to $r$, and 
\begin{eqnarray}
\Pi_{s}(r) & = & 1+\frac{\left(f''(r)-\frac{f'(r)}{r}\right)\left(J_{z}-Hs\right)^{2}s^{2}}{2r^{2}\Sigma_{s}(r)^{3}}\,,\\
\Sigma_{s}(r) & = & 1-\frac{f'(r)}{2r}s^{2}\,,\\
R_{s}(r) & = & \left(H-\frac{f'(r)}{2r}sJ_{z}\right)^{2}\\
 & - & f(r)\left\{ \Sigma_{s}(r)^{2}+\frac{\left(J_{z}-Hs\right)^{2}}{r^{2}}\right\} \,.\nonumber 
\end{eqnarray}
Again we are interested in bound motion, so we require two turning
points $r_{a}$ and $r_{p}$ where $dr/d\tau=0$ which corresponds
to $R_{s}(r)=0$. We rescale, as before, $r\mapsto rM$, $J_{z}\mapsto J_{z}M$,
$\Lambda\mapsto\Lambda M^{-2}$ and now also $s\mapsto sM$. Note
that for test particle motion our dimensionless spin parameter $s$
necessarily satisfies the condition $-1<s<1$, see \cite{Hartl03a}.
Moreover, we substitute 
\begin{equation}
L=J_{z}-Hs\label{eq:defL}
\end{equation}
for mathematical convenience. Then $R_{s}(r)$ can be rewritten as
\begin{equation}
R_{s}(r)=\frac{\Lambda\left(1+\frac{\Lambda}{3}s^{2}\right)^{2}}{3r^{7}}P_{9}(r)\label{eq:RP9}
\end{equation}
where 
\begin{equation}
P_{9}(r)=r^{9}+ar^{7}+br^{6}+cr^{5}+dr^{4}+er^{3}+gr+h\label{eq:P9}
\end{equation}
with 
\begin{eqnarray}
a & = & \frac{\Big(H\left(1+\frac{\Lambda}{3}s^{2}\right)+\frac{\Lambda}{3}Ls\Big)^{2}+\frac{\Lambda}{3}L^{2}}{\frac{\Lambda}{3}\,\left(1+\frac{\Lambda}{3}s^{2}\right)^{2}}-\dfrac{3}{\Lambda}\,,\\
b & = & \frac{2}{\frac{\Lambda}{3}\,\left(1+\frac{\Lambda}{3}s^{2}\right)}\,,\\
c & = & -\,\frac{\, L^{2}}{\frac{\Lambda}{3}\,\left(1+\frac{\Lambda}{3}s^{2}\right)^{2}}\,,\\
d & = & -\,\frac{2\left(\left(L+Hs\right)^{2}-s^{2}\right)}{\frac{\Lambda}{3}\,\left(1+\frac{\Lambda}{3}s^{2}\right)}\,+\frac{2L\left(2L+Hs\right)}{\frac{\Lambda}{3}\left(1+\frac{\Lambda}{3}s^{2}\right)^{2}}\,,\\
e & = & -\,\frac{4s^{2}\left(1+\frac{\Lambda}{4}s^{2}\right)}{\frac{\Lambda}{3}\left(1+\frac{\Lambda}{3}s^{2}\right)^{2}}\,,\\
g & = & \frac{s^{2}\left(\left(L+Hs\right)^{2}-s^{2}\right)}{\frac{\Lambda}{3}\,\left(1+\frac{\Lambda}{3}s^{2}\right)^{2}}\,,\\
h & = & \frac{2\, s^{4}}{\frac{\Lambda}{3}\,\left(1+\frac{\Lambda}{3}s^{2}\right)^{2}}\,.
\end{eqnarray}
The number of zeros of $P_{9}(r)$ determines the types of motion
possible in the corresponding spacetime. Finding the maximum number
of positive real zeros in this case is not so easy, though, since
the signs of $a$ and $d$ are unclear. (The sign of $g$ does not
matter because the signs of $e$ and $h$ are already different.)
In any case, from Descartes' rule of signs we find that the number
of positive real zeros can only be $0,$ $2,$ $4$ or $6$. Again,
in order to consider bound motion we must have at least four positive
real zeros. We argue that in this case \eqref{eq:P9} can be written
as 
\begin{equation}
\begin{split}P_{9}\left(r\right)=(r-r_{\gamma})(r-r_{\beta})(r-r_{\alpha})(r-r_{1})\\
\times(r-r_{p})(r-r_{a})(r-r_{2})(r-r_{I})(r-\bar{r}_{I})\label{eq:VrS}
\end{split}
\end{equation}
with $r_{\gamma}<r_{\beta}<r_{\alpha}<0<r_{1}<r_{p}<r_{a}<r_{2}$
and $r_{I}$, $\bar{r}_{I}$ being some non-real complex zero and
its conjugate, respectively. The knowledge of a non-real zero was
obtained by checking the signs of the coefficients in the relevant
parameter space. To exclude the case with six positive real zeros,
we observe that in the relevant range of parameters, $0<\Lambda<1/9$
and $-1<s<1$, only three mergers of zeros exist. Therefore we conclude
to have one pair of complex zeroes.

Bound motion exists only between $r_{p}$ and $r_{a}$. Since also
for the spinning particle the motion is fully described by the parameters
$r_{a}$ and $r_{p}$, we find expressions for $H(r_{a},r_{p})$ and
$L(r_{a},r_{p})$ by setting $R_{s}(r_{a})=0$ and $R_{s}(r_{p})=0.$
With the help of \eqref{eq:rarppe} the expressions are converted
to $H(p,e)$ and $L(p,e)$. The next task is to find the boundaries
of the region of bound motion in the $(p,e)$-plane, i.e. the separatrices.
Again, the boundaries are given by the merging of two zeros, $r_{1}=r_{p}$
or/and $r_{p}=r_{a}$ or/and $r_{a}=r_{2}$. Unfortunately, the explicit
expressions for the zeros are not as easily found as in the non-spinning
case. However, there is a way around: We simply exploit the fact that
the values of $H$ and $L$ for the homoclinic orbits are identical
to the ones for the unstable circular orbits which they approach asymptotically.
Having access to the values of $H$ and $L$ for circular orbits by
solving $R_{s}(r)=0$ and $R_{s}'(r)=0$, the only computation needed
is to find the intersection point of the lines of constant $H$ and
constant $L$ in the $(p,e)$-plane. In this way we obtain the values
for $p$ and $e$ of the homoclinic orbits corresponding to the separatrices.

\begin{figure}[tph]
\center{\includegraphics[width=7cm]{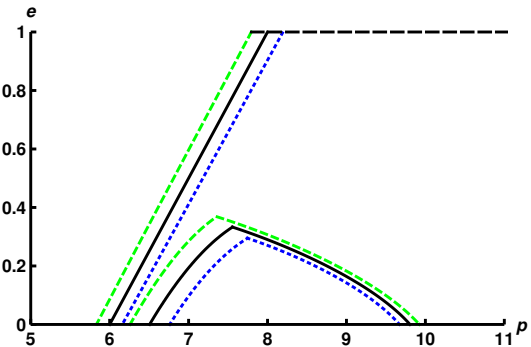}}
\caption{The figure presents the regions of bound motion in the $(p,e)$-plane
for different values of $s$ for $\Lambda=0$ as well as $\Lambda=0.0005$.
The straight lines on the left correspond to $\Lambda=0$ and the
triangles on the right correspond to $\Lambda=0.0005$. In either
case, the (green) dashed line corresponds to $s=0.1$, the black solid
line to $s=0$, and the (blue) dotted line to $s=-0.1$.}

\label{Fig:SepS} 
\end{figure}

In Fig. \ref{Fig:SepS} the region of bound motion is shown in a $(p,e)$-diagram
for both a spinning and a non-spinning particle moving in Schwarzschild
and Schwarzschild-de-Sitter spacetime. Here, we have fixed the cosmological
constant either to $\Lambda=0$ or to $\Lambda=0.0005$ and varied
the value of the spin parameter.

From Fig. \ref{Fig:SepS} we read that for a spinning particle in
Schwarzschild spacetime the region of bound motion is infinitely large
as it is for geodesic motion. The well-known shift of the ISCO due
to the spin is visible, such that for positive spin it is moved inwards
and for negative spin outwards, cf. \cite{Tanaka96} and also \cite{Suzuki98,Favata11}.
This reflects the coupling of the particle's spin to its orbital angular
momentum. If they are parallel to each other the resulting force is
repulsive, while it is attractive if they are antiparallel \cite{Tanaka96}.
Notice that the upper boundary is given by $e=1$ which corresponds
to parabolic orbits and marks the transition from bound motion to
unbound orbits. By and large, the general shape of the separatrices
resembles the one for non-spinning particles in Schwarzschild spacetime.

\begin{figure}[tph]
\center{\includegraphics[width=7cm]{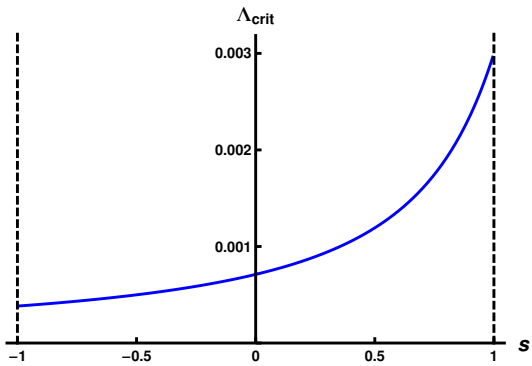}} \caption{The figure presents the dependence of the critical value of $\Lambda$
on the spin parameter. It monotonically increases with the spin. The
boundaries $s=-1$ and $s=1$ are due to physical restrictions. }

\label{Fig:Lambdacrit} 
\end{figure}

When a positive cosmological constant is considered, the value of
the maximal eccentricity becomes smaller for negative and larger for
positive spin. Correspondingly, the critical value of $\Lambda$ is
also shifted: for positive spin $\Lambda_{\text{crit}}$ is bigger
than for the spinless case and for negative spin it is smaller. The
dependence of $\Lambda_{\text{crit }}$ on the spin is shown in Fig.~\ref{Fig:Lambdacrit}.

It reveals that for $\Lambda<\Lambda_{\text{crit}}(s=0)=4/5625$ bound
motion is possible for all positive spin values but not for all negative
spin values. This is the reason why for our particular choice of $\Lambda=0.0005$
it is not possible to have bound orbits with spins that are smaller
than $\approx-0.5$. As soon as the chosen value of $\Lambda$ drops
below $\Lambda_{\mathrm{crit}}(s=-1)\approx0.0004$ bound motion is
possible for all spin values $-1<s<1$.

The influence of the spin on the ISCO in the Schwarzschild-de Sitter
spacetime is similar as in the Schwarzschild spacetime. Again we see
that for positive spin parameters the ISCO gets shifted towards the
center and for negative spin parameters away from the center. For
the OSCO it is the other way around. By taking into account these
two characteristics as well as the shift of the maximal eccentricity,
it can be immediately seen that the region of bound orbits becomes
smaller for negative spin and gets larger for positive spin.

One might think that a large positive spin is somehow able to destroy
the existence of the heteroclinic orbit sitting at maximal eccentricity.
However, even if the spin is chosen to have its maximal value of $1$
the shape of the region of bound motion does not change. The triangle
survives and with it the heteroclinic orbit. If $\Lambda$ approaches
zero the maximal eccentricity goes to $1$ and the OSCO to infinity,
for any spin value. In this case the separatrix resembles the one
of Schwarzschild, only shifted closer to the center for $s>0$ and
farther away from the center for $s<0$.

In the following section we change from the parametrization of bound
orbits in terms of $p$ and $e$ to the frequency domain. This will
allow us to discuss the phenomenon of isofrequency pairing.

\section{Isofrequency Pairing }

As mentioned before, Barack and Sago \cite{Barack11} observed that
for bound orbits of spinless test-particles in the Schwarzschild spacetime
the transformation from the constants of motion $H$ and $J_{z}$
to the radial and azimuthal frequencies becomes degenerate in the
highly relativistic regime. In particular, it was demonstrated that
orbits exist with different eccentricities but with the same frequency
pair.

Here, we show how the picture of this degeneracy phenomenon changes
if a cosmological constant is turned on and if the particle's spin
is incorporated. Generally, the frequencies are defined as \cite{Cutler94}
\begin{equation}
\Omega_{r}=\frac{2\pi}{T_{r}}\,,\qquad\Omega_{\phi}=\frac{\Delta\phi}{T_{r}}\label{eq:OmegaPhi}
\end{equation}
with 
\begin{eqnarray}
T_{r}=2\int_{r_{p}}^{r_{a}}\frac{dt}{dr}dr=2\int_{r_{p}}^{r_{a}}\frac{dt}{d\tau}\left(\frac{dr}{d\tau}\right)^{-1}dr\,,\label{eq:Tr}\\
\Delta\phi=2\int_{r_{p}}^{r_{a}}\frac{d\phi}{d\tau}\left(\frac{dr}{d\tau}\right)^{-1}dr\,.\qquad\quad\label{eq:DeltaPhi}
\end{eqnarray}
In correspondence with our choice of units in the previous sections
we rescale $\Omega_{r}\rightarrow\Omega_{r}/M$ and analogously $\Omega_{\phi}\rightarrow\Omega_{\phi}/M$.

\subsection{Non-spinning particles}

Again, we start with the geodesic motion. From equations \eqref{eq:teom}-\eqref{eq:Vr}
we obtain 
\begin{align}
T_{r}= & 2H\sqrt{\frac{3}{\Lambda}}\int_{r_{p}}^{r_{a}}\frac{r^{2}dr}{f(r)\sqrt{rP_{5}(r)}}\,,\label{eq:TrVr}\\
\Delta\phi= & 2L\sqrt{\frac{3}{\Lambda}}\int_{r_{p}}^{r_{a}}\frac{dr}{\sqrt{rP_{5}(r)}}\,,\label{eq:DeltaPhiVr}
\end{align}
with $f(r)$ from (\ref{eq:f}) and $P_{5}(r)$ from (\ref{eq:P5}).
As the polynomial $rP_{5}\left(r\right)$ under the root in the denominator
of the integrand is of order $6$, the integral is of hyperelliptic
type, which cannot be integrated in terms of elementary functions.

If we use, as before, the $(p,e)$ parametrization of the bound orbits,
we may substitute the integration variable $r$ in \eqref{eq:TrVr}
and \eqref{eq:DeltaPhiVr} according to 
\begin{equation}
r=\frac{p}{1+e\cos\chi}
\end{equation}
where the new integration variable $\chi$ is the relativistic anomaly.
Then the boundary values of the integral change to $0$ and $\pi$.

Since we want to compare the frequencies of different orbits, we choose
in analogy to Warburton et al. \cite{Warburton13} the $(\Omega_{\phi},e)$
parametrization for our analysis of bound orbits, which we are allowed
to because $\Omega_{\phi}$ monotonically decreases with $p$ if the
eccentricity is held fixed. In order to do this we have to deal with
several obstacles. First, we cannot analytically invert the $\Omega_{\phi}$
integral to obtain $p$ as a function of $e$ and $\Omega_{\phi}$.
To circumvent this hindrance the value for $e$ is fixed in the integral
for the frequency, so that the value for $p$ can be computed using
a root-finding method for any allowed $\Omega_{\phi}$. In this way
we obtain the value for $p$ for any given $e$ and $\Omega_{\phi}$;
that is to say, we numerically acquire a function $p(e,\Omega_{\phi})$.
Hence, the radial frequency can also be written as a function $\Omega_{r}(e,\Omega_{\phi})$
and we are able to plot contour lines for constant $\Omega_{r}$ into
an $(\Omega_{\phi},e)$-diagram.

Secondly, we encounter problems close to the separatrices. Both $T_{r}$
and $\Delta\phi$ diverge at the separatrices making it numerically
challenging to perform the computations for the frequencies. Luckily,
there exists an approximation scheme for hyperelliptic integrals developed
by Sochnev~\cite{Sochnev68} based on the approximation of irrational
numbers. We rewrite the integrals of \eqref{eq:TrVr} and \eqref{eq:DeltaPhiVr}
with the substitution 
\begin{equation}
r=\frac{(r_{a}+r_{p})+(r_{a}-r_{p})x}{2}\label{eq:SochnevSubs}
\end{equation}
to obtain 
\begin{align}
T_{r}= & A(r_{p},r_{a},\Lambda)\int_{-1}^{1}\frac{V_{t}(x)dx}{\sqrt{V_{r}(x)}}\,,\label{eq:TrSochnev}\\
\Delta\phi= & B(r_{p},r_{a},\Lambda)\int_{-1}^{1}\frac{dx}{\sqrt{V_{r}(x)}}\,,\label{eq:DeltaPhiSochnev}
\end{align}
where $V_{t}(x)$ is a rational function whose denominator has no
zeros in the integration interval and 
\begin{equation}
V_{r}(x)=(1-x^{2})(1+k_{1}x)(1+k_{2}x)(1+k_{3}x)(1+k_{4}x)\label{eq:Vrx}
\end{equation}
with 
\begin{eqnarray*}
k_{1} & = & \frac{r_{a}-r_{p}}{r_{a}+r_{p}}\,,\\
k_{2} & = & \frac{r_{a}-r_{p}}{(r_{a}-r_{0})+(r_{p}-r_{0})}\,,\\
k_{3} & = & \frac{r_{a}-r_{p}}{(r_{a}-r_{1})+(r_{p}-r_{1})}\,,\\
k_{4} & = & -\frac{r_{a}-r_{p}}{(r_{2}-r_{a})+(r_{2}-r_{p})}\,,
\end{eqnarray*}
which satisfy $0\leq k_{2}\leq k_{1}\leq k_{3}\leq1$ and $-1\leq k_{4}\leq0$.
If we get close to the separatrices either $k_{3}$ or $-k_{4}$ approaches
$1$. For each of these two cases Sochnev's method allows to approximate
the integrals (\ref{eq:TrSochnev}) and (\ref{eq:DeltaPhiSochnev})
in terms of elementary integrals, see the Appendix. Consequently,
the frequencies close to the separatrix can be approximately expressed
in terms of elementary functions. Now we have the tools we need for
analyzing the behavior of the frequencies in the region of bound motion.

\begin{figure}[tph]
\center{\includegraphics[width=8cm]{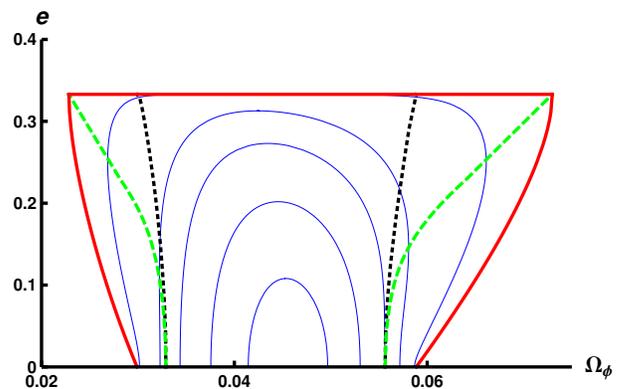}}
\caption{The figure depicts the phenomenon of isofrequency pairing for bound
orbits in the $(\Omega_{\phi},e)$-plane for a fixed $\Lambda=0.0005$.
The thick (red) boundary lines correspond to the separatrices and
confine the region of bound motion. The (blue) solid lines inside
this region coorespond to constant values of $\Omega_{r}$. The (green)
dashed lines are the singular curves, i.e. the locus where the Jacobian
determinant of the transformation from $(p,e)$ to $(\Omega_{r},\Omega_{\phi})$
vanishes. Isofrequency pairing occurs in two domains each of which
is bounded by a separatrix and a curve of Circular Orbit Duals (COD).
The COD curves are shown by the dotted black lines. }

\label{Fig:ISFPwithoutSpin} 
\end{figure}

In Fig. \ref{Fig:ISFPwithoutSpin} the region of bound orbits is shown
in the $(\Omega_{\phi},e)$-plane. It is the shape of this region
that is striking. It looks no longer like a triangle, as in the $(p,e)$-representation,
but more like a trapezoid. The tip of the triangle is stretched out
to a straight line at $e_{\text{max}}$. In the $(p,e)$-representation
the heteroclinic orbit corresponds to a single point -- the tip of
the triangle -- since it has a uniquely defined pair $(r_{p},r_{a})$.
By contrast, the azimuthal frequency is not uniquely defined for the
heteroclinic orbit. Note that the original definition of the frequencies
is valid only within the region of bound orbits. On the boundaries,
that is at the separatrices, the frequencies are defined only by a
continuous extension which assigns unique frequencies to the homoclinic
orbits. However, in the case of the heteroclinic orbit the value of
$\Omega_{\phi}$ depends on how the orbit is approached. This is the
reason why the heteroclinic orbit is stretched out to a straight line
in the $(\Omega_{\phi},e)$-diagram. A bound orbit that lies close
to this line could be called a ``whirl-whirl orbit'' because it
periodically changes between a large number of whirls near its apastron
and a large number of whirls near its periastron.


The main property we are interested in is the isofrequency pairing.
This phenomenon is easily seen in the diagram close to the two separatrices.
Looking at a contour line for $\Omega_{r}=\mathrm{constant}$ near
one of the separatrices, we see that it has two intersection points
with a contour line for $\Omega_{\phi}=\mathrm{constant}$, i.e.,
with a vertical line in this diagram. Since these intersection points
correspond to orbits of different eccentricities, we conclude that
there exist two geometrically distinct orbits with the same pair of
frequencies. In mathematical terms this means that the transformation
from the frequencies $\left(\Omega_{r},\Omega_{\phi}\right)$ to $\left(p,e\right)$
is not one-to-one. In order to prove this degeneracy it is sufficient
to show that the Jacobi determinant 
\begin{equation}
J=\Bigg\vert\frac{\partial\left(\Omega_{r},\Omega_{\phi}\right)}{\partial\left(p,e\right)}\Bigg\vert
\end{equation}
becomes singular somewhere within the region of bound orbits. In Fig.~\ref{Fig:ISFPwithoutSpin}
these singular points can be found as the points where the tangents
to the contour lines of $\Omega_{r}=\text{constant}$ become vertical.
This happens along the two (green) dashed curves in Fig.~\ref{Fig:ISFPwithoutSpin}
which are called the \emph{singular curves}. To verify that $J$ does
have two zeros close to $e=0$, one may perform a Taylor expansion
of $J$ about $e=0$ up to first order, \begin{widetext} 
\begin{eqnarray*}
J & = & \frac{eP_{10}\left(p\right)}{4p^{4}\sqrt{p\left(\frac{\Lambda}{3}p^{3}-1\right)}\left(\frac{\Lambda}{3}p^{3}-p+2\right)\left(\frac{\Lambda}{3}(4p-15)p^{3}-p+9\right)^{3/2}}+O(e^{2})
\end{eqnarray*}
with 
\[
P_{10}\left(p\right)=15\Lambda^{3}p^{10}-50\Lambda^{3}p^{9}+30\Lambda^{2}p^{8}-315\Lambda^{2}p^{7}+9\Lambda(80\Lambda-1)p^{6}-45\Lambda p^{5}+918\Lambda p^{4}-2241\Lambda p^{3}+108p^{2}-1053p+2322\,.
\]
\end{widetext} Numerically one finds that the tenth order polynomial
$P_{10}(p)$ has precisely two positive real zeros lying withing the
allowed range of $p$ values for bound motion, for all $0<\Lambda<1/9$.

The isofrequency pairs lie on opposite sides of one of the singular
curves, i.e., each orbit that is located between a separatrix and
a singular curve has a partner orbit on the other side of the corresponding
singular curve which is geometrically different but has the same pair
of frequencies. The regions where isofrequency pairing occurs are
bounded by the socalled ``Circular Orbit Dual'' (COD) curves which
are represented by the black dashed lines. A point on a COD curve
corresponds to an orbit that has the same frequencies as a stable
circular orbit situated between one of the singular curves and the
corresponding separatrix.

In comparison to the Schwarzschild spacetime, the most interesting
new feature of isofrequency pairing in the Schwarzschild-de Sitter
spacetime is in the fact that this phenomenon now occurs not only
in the highly relativistic regime close to the center but also in
a region far away from the center.

\begin{figure}[tph]
\center{\includegraphics[width=0.4\textwidth]{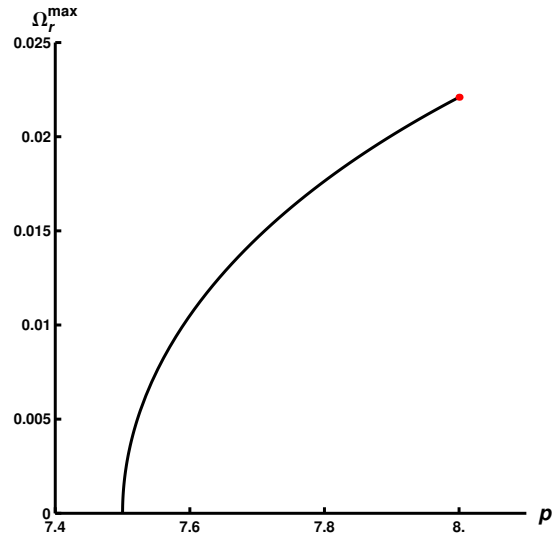}}
\caption{The figure shows how the value of $\Omega_{r}^{\text{max}}$ and the
position where this value is taken depend on $\Lambda$. The (red)
dot marks the Schwarzschild case with $\Lambda=0$. Along the solid
black line $\Lambda$ increases up to its maximal value of 1/9 where
the bound orbits vanish. }

\label{Fig:Omegarmaxp} 
\end{figure}

While we could read from Fig. \ref{Fig:Seppe} how the ISCO radius,
the OSCO radius and the maximal eccentricity depend on $\Lambda$,
Fig.~\ref{Fig:ISFPwithoutSpin} gives us information on the frequencies
of bound motion. A quite interesting property is the existence of
a maximal radial frequency. From Fig.~\ref{Fig:ISFPwithoutSpin}
we read that the lines of constant $\Omega_{r}$ have the shape of
semicircles which become smaller in the center of the diagram. The
value of $\Omega_{r}$ varies between $\Omega_{r}=0$ on the separatrices
and a maximal value when the semicircle is contracted to just one
single point. Fig. \ref{Fig:Omegarmaxp} shows a parametric plot of
the maximal value of $\Omega_{r}$ against $p$, where the parameter
is the cosmological constant. As $\Omega_{r}$ takes its maximal value
on the axis $e=0$, i.e., for the limiting case of a circular orbit,
$p$ is the same as the radius coordinate. We see that the position
where $\Omega_{r}$ takes its maximal value varies between $p=8$
for $\Lambda=0$ and $p=7.5$ for $\Lambda=\Lambda_{\text{crit}}=4/5625$.

Moreover, we can read from Fig.~\ref{Fig:Omegarmaxp} that the presence
of a positive cosmological constant introduces another kind of degeneracy.
Notice that $\Omega_{r}^{\text{max}}$ decreases slightly when $\Lambda$
is slightly increased. Transferring this change to Fig.~\ref{Fig:ISFPwithoutSpin}
we can think of the lines of constant $\Omega_{r}$ getting shifted
towards the center. Imagine that we have a frequency pair close to
the right separatrix. Choose the outer contour line of $\Omega_{r}=\text{constant}$
and let some vertical line of constant $\Omega_{\phi}$ intersect
it twice and mark these two points. As described above, they are two
orbits with the same frequencies. If the cosmological constant is
changed a little bit the line of constant $\Omega_{r}$ gets shifted
either to the right if $\Lambda$ is reduced or to the left if $\Lambda$
is amplified. However, we would still have two intersection points
with the vertical line representing the azimuthal frequency we have
fixed before. If $\Lambda$ is reduced the two intersection points
diverge while for amplified $\Lambda$ the intersection points approach
one another, i.e. their eccentricities change. This means that infinitely
many physically distinct pairs of orbits have the same frequencies
as the originally fixed one if we allow the cosmological constant
to take values in a certain interval.

In the next section we discuss what happens to the degeneracy features
when a spin of the particle is added.

\subsection{Spinning particles in the equatorial plane}

As before, we are interested in the influence of a particle's spin
on its motion and dynamical properties. From \eqref{eq:teomS}-\eqref{eq:phieomS}
we obtain 
\begin{align}
T_{r}= & 2\sqrt{\frac{3}{\Lambda}}\int_{r_{p}}^{r_{a}}\frac{\tilde{V}{}_{t}^{s}(r)}{\sqrt{rP_{9}\left(r\right)}}dr\,,\label{eq:TrS}\\
\Delta\phi= & 2L\sqrt{\frac{3}{\Lambda}}\int_{r_{p}}^{r_{a}}\frac{\tilde{V}{}_{\phi}^{s}(r)}{\sqrt{rP_{9}\left(r\right)}}dr\,,\label{eq:DeltaPhiS}\\
\end{align}
with $P_{9}(r)$ from \eqref{eq:P9} and 
\begin{align}
\tilde{V}{}_{t}^{s}(r)= & \frac{r^{2}\left[H\left(r^{3}+\left(\frac{\Lambda}{3}r^{3}-1\right)s^{2}\right)+\left(\frac{\Lambda}{3}r^{3}-1\right)sL\right]}{\left(-\frac{\Lambda}{3}r^{3}+r-2\right)\left(1+\frac{\Lambda}{3}s^{2}\right)}\,,\\
\tilde{V}{}_{\phi}^{s}(r)= & \frac{r^{2}\left(r^{3}+\left(\frac{\Lambda}{3}r^{3}+2\right)s^{2}\right)}{\left(r^{3}+\left(\frac{\Lambda}{3}r^{3}-1\right)s^{2}\right)\left(1+\frac{\Lambda}{3}s^{2}\right)}\,.
\end{align}
The order of the polynomial $rP_{9}\left(r\right)$ under the root
in the denominator of the integrand is $10$. Therefore we now have
a considerably more difficult kind of hyperelliptic integral than
in the spinless case.

Following the same procedure as in the non-spinning case, we transform
the integrals with $r=\frac{p}{1+e\cos\chi}$ to obtain the resulting
frequencies as functions of $(p,e)$. Therewith, we are able to reparametrize
the radial frequency, again, as a function of $(\Omega_{\phi},e)$
so that the comparison of different orbits and their frequencies is
simplified. Since the inclusion of the spin does not make the system
easier to be solved, we encounter the same numerical problems as before,
i.e. the numerical inversion to $p(\Omega_{\phi},e)$ and the divergencies
close to the separatrices, as we have in the non-spinning case. Luckily,
these problems can be tackled also by the same method, only the expressions
become more complicated.

Employing Sochnev's approach of approximating hyperelliptic integrals
we rewrite the integrals of \eqref{eq:TrS} and \eqref{eq:DeltaPhiS}
with the substitution given in \eqref{eq:SochnevSubs} to obtain 
\begin{align}
T_{r}= & A^{s}(r_{p},r_{a},\Lambda,s)\int_{-1}^{1}\frac{V_{t}^{s}(x)}{\sqrt{V_{r}^{s}(x)}}dx\,,\label{eq:TrSpin}\\
\Delta\phi= & B^{s}(r_{p},r_{a},\Lambda,s)\int_{-1}^{1}\frac{V_{\phi}^{s}(x)}{\sqrt{V_{r}^{s}(x)}}dx\,,\label{eq:DeltaPhiSpin}
\end{align}
where $V_{t}^{s}(x)$ and $V_{\phi}^{s}(x)$ are rational functions
whose denominators have no zeros in the integration interval and 
\begin{align}
V_{r}^{s}(x)=(1-x^{2})(1+k_{1}x)(1+k_{2}x)(1+k_{3}x)(1+k_{4}x)\label{eq:Vrsx}\\
\times(1+k_{5}x)(1+k_{6}x)\left(x-\left(a_{I}+ib_{I}\right)\right)\left(x-\left(a_{I}-ib_{I}\right)\right)
\end{align}
with

\begin{eqnarray*}
k_{1} & = & \frac{r_{a}-r_{p}}{r_{a}+r_{p}}\,,\\
k_{2} & = & \frac{r_{a}-r_{p}}{(r_{a}-r_{1})+(r_{p}-r_{1})}\,,\\
k_{3} & =- & \frac{r_{a}-r_{p}}{(r_{2}-r_{a})+(r_{2}-p_{2})}\,,\\
k_{4} & = & \frac{r_{a}-r_{p}}{(r_{a}-r_{\alpha})+(r_{p}-r_{\alpha})}\,,\\
k_{5} & = & \frac{r_{a}-r_{p}}{(r_{a}-r_{\beta})+(r_{p}-r_{\beta})}\,,\\
k_{6} & = & \frac{r_{a}-r_{p}}{(r_{a}-r_{\gamma})+(r_{p}-r_{\gamma})}\,,
\end{eqnarray*}
which satisfy $0\leq k_{6}\leq k_{5}\leq k_{4}\leq k_{1}\leq k_{2}\leq1$
and $-1\leq k_{3}\leq0$. As in the non-spinning case, we use Sochnev's
method for evaluating these integrals near the separatrices, see the
Appendix.

In Fig. \ref{Fig:ISFPSpin} plots of the isofrequency pairing phenomenon
for particles with two different spin values $(0.1,-0.1)$ moving
in the same Schwarzschild-de Sitter spacetime with $\Lambda=0.0005$
are shown. The qualitative shape of the region of bound motion is
similar to that of the non-spinning particle. The quantitative differences
become apparent in the characteristic features of bound motion and
its boundaries. Some basic facts that we have already seen in the
$(p,e)$-diagram are obvious, such as the shifts of the maximal eccentricity,
of the ISCO and of the OSCO. However, what interests us is the impact
on the domain where isofrequency pairing occurs and the question of
whether further degeneracies due to the spin emerge.

\begin{figure}[tph]
\centerline{\includegraphics[width=0.45\textwidth]{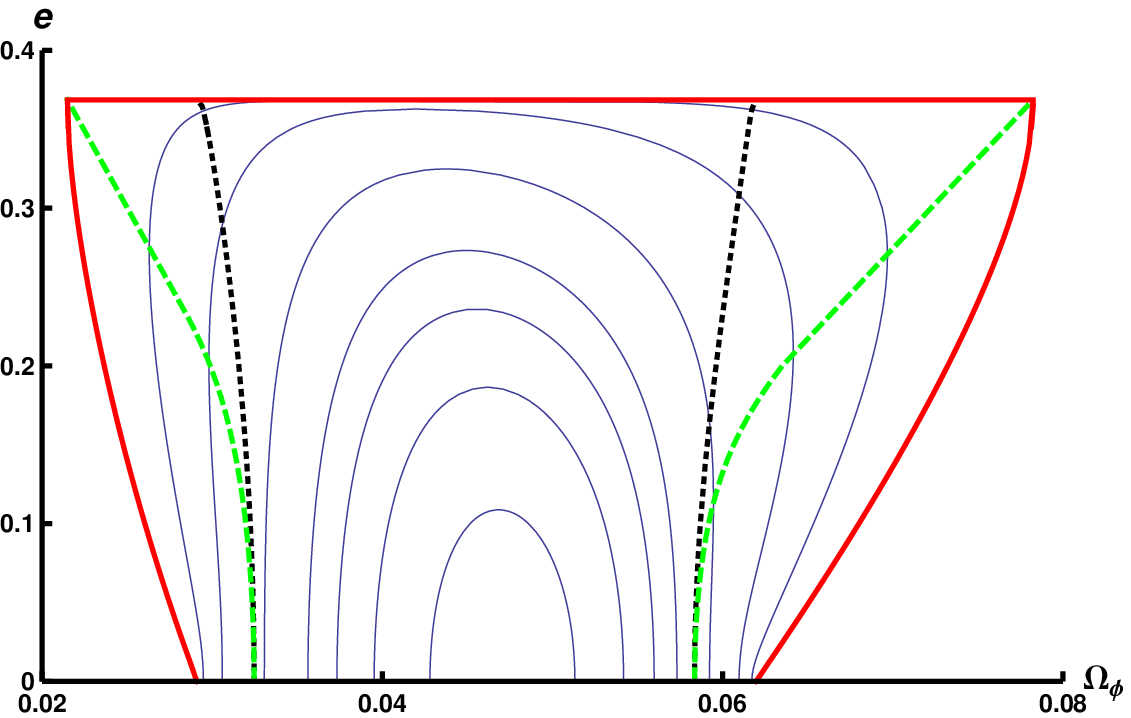}}
\vspace{0.5cm}
 \centerline{\includegraphics[width=0.45\textwidth]{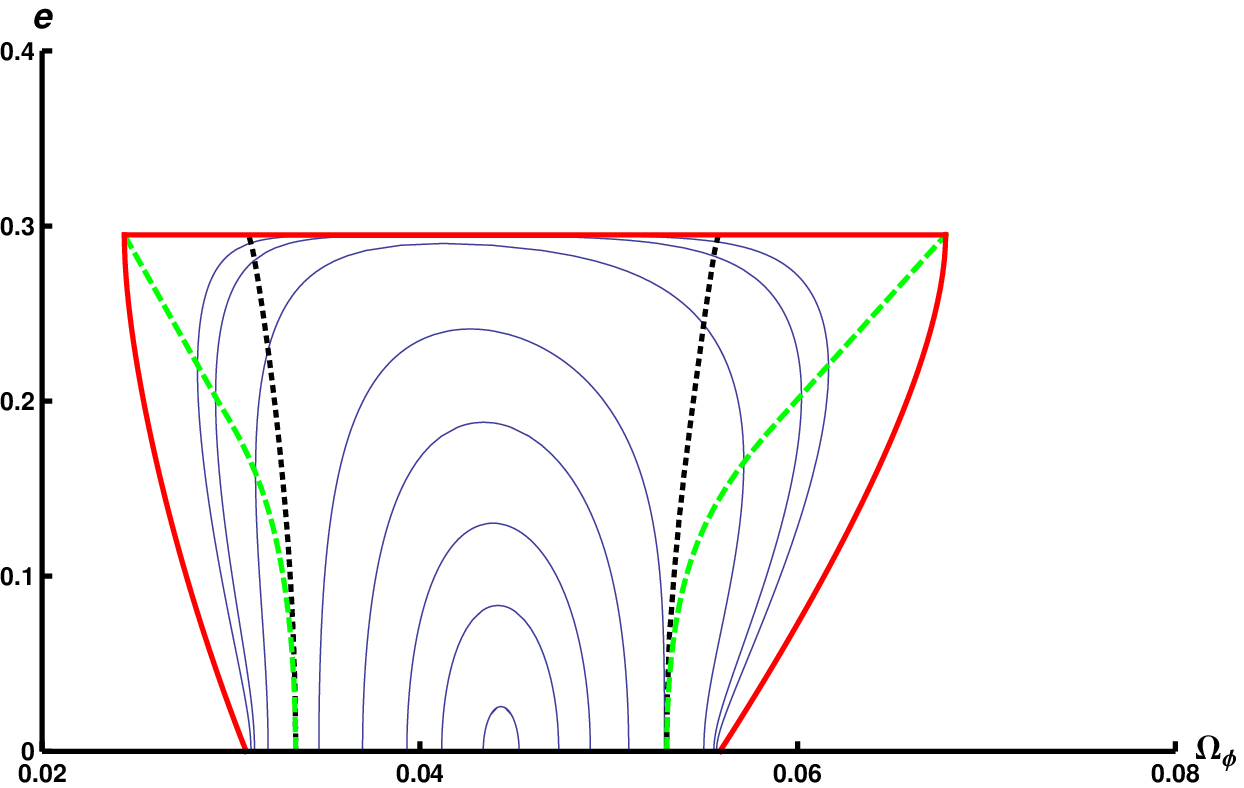}
}\caption{The figure depicts the phenomenon of isofrequency pairing for bound
orbits in the $(\Omega_{\phi},e)$-plane for a fixed $\Lambda=0.0005$.
The upper figure shows the chracteristics for $s=0.1$ and the lower
one corresponds to $s=-0.1$. The thick (red) boundary lines correspond
to the separatrices and confine the region of bound motion. The (blue)
solid lines inside these regions coorespond to constant values of
$\Omega_{r}$. The (green) dashed line represents the singular curve,
i.e. the locus where the Jacobian determinant of the transformation
from $(p,e)$ to $(\Omega_{r},\Omega_{\phi})$ vanishes. The Circular
Orbit Duals (COD) marking the boundaries of the domains where isofrequency
pairing occur are shown by the black dotted lines.}

\label{Fig:ISFPSpin} 
\end{figure}

First, we qualitatively compare the size of the region where isofrequent
orbits exist. It can be characterized by the azimuthal frequency at
the ISCO (OSCO) and the one at the intersection of the COD line with
the horizontal axis. Only orbits having an azimuthal frequency within
this range do have isofrequent partners. The size of the allowed frequency
interval decreases if the spin is chosen to be negative and increases
if the spin value is positive. Although the region shrinks for negative
spin it will never completely vanish as long as bound motion exists.
Therefore, the spin does not destroy this degeneracy in the fundamental
frequencies of the orbital motion.

\begin{figure}[tph]
\center{\includegraphics[width=8cm]{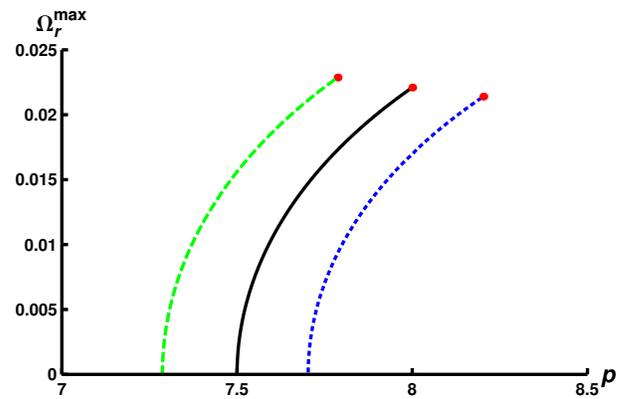}}
\caption{The figure shows the change of $\Omega_{r}^{\max}$ in value as well
as in position from the center when the value of $\Lambda$ is varied.
The (red) dots mark the Schwarzschild case with $\Lambda=0$. Along
each curve $\Lambda$ increases up to its maximal value when $\Omega_{r}^{\text{max}}$
vanishes. The black solid curve correponds to $s=0$, the (green)
dashed one to $s=0.1$ and the (blue) dotted one to $s=-0.1$.}

\label{Fig:OmegarmaxSpin} 
\end{figure}

Next, we compare the evolution of the maximal radial frequency for
different spin values in Fig. \ref{Fig:OmegarmaxSpin}. We notice
immediately the shift of the entire curve closer to the center for
positive spin and further away for negative spin values. This coincides
with the trend of the shifts of the ISCO and OSCO. We also see the
spin dependence of the value for $\Omega_{r}^{\text{max}}$ for $\Lambda=0$
(bold dots). More generally, if we choose a value for $\Omega_{r}^{\text{max}}$
on the vertical axis and determine the intersection points with the
three curves not only the position differs but also the corresponding
$\Lambda$ is not the same for the different systems. 

We will now investigate if the spin induces a further degeneracy in
the sense that isofrequency pairs with different spin values may have
the same frequencies. 
We can deduce this from Fig.~\ref{Fig:ISFPSpin}. Choose a line of
constant radial frequency that lies in the isofrequency range. Then
draw a line of constant azimuthal frequency that corresponds to a
circular orbit located outside the range between the ISCO and the
OSCO and mark the two intersection points. We know from the analysis
of $\Omega_{r}^{\text{ma}x}$ that the maximal radial frequency increases
with positive spin and decreases with negative spin, keeping the cosmological
constant fixed. Now, imagine the line of constant $\Omega_{r}$ shrinking
and spreading slowly for small negative and postive values for the
spin parameter, respectively. The two intersection points will diverge
for positive spin and approach one another for negative spin leading
to six different orbits that have different eccentricities. For example,
if the inital intersection points belong to a non-spinning particle
and the spin is slightly changed to positive or negative values, there
exist infinitely many physically distinct pairs of orbits having the
same frequencies but different spin values. Even the spin direction
is different in this example. If we allow both the cosmological constant
and the spin parameter to vary we get two-parameter families of isofrequency
pairs with the same frequecies, parametrized by $(\Lambda,s)$. 

\section{Conclusions and outlook}

In this work we investigated the characteristics of isofrequency pairing
for both geodesic motion of a test-particle and the non-geodesic motion
of a spinning test-particle moving in the equatorial plane of Schwarzschild-de
Sitter spacetime. In contrast to the case without a cosmological constant,
there exist \textit{two} regions in the domain of bound orbits where
isofrequency pairing occurs. More precisely, it is not only the strong
field regime that exhibits such a feature but also a region close
to the outermost stable circular orbit. This is associated with a
greater variety of zoom-whirl orbits than in the case without a cosmological
constant: There are now not only orbits that whirl near the periastron
but also orbits that whirl near the apastron.

Generally, adding a cosmological constant and/or the spin leads to
the emergence of additional degeneracies in the frequencies. This
occurs already for arbitrarily small values of the cosmological constant
and the spin. At least in principle, this additional degeneracy is
of relevance in view of gravitational wave data analysis. Here we
may think of an EMRI which, among all possible gravitational wave
sources, is the closest physical realization of the dynamical system
considered in this paper. Whenever isofrequency pairing occurs, knowledge
of the fundamental frequencies alone does not determine the shape
of the orbit, i.e., additional information on the spectrum has to
be taken into account.



The most obvious plan for follow-up work would be to consider more
general spacetimes, such as the Kerr-deSitter-NUT... spacetime. In
addition, there are several other avenues for future studies on isofrequency
pairing which we would like to mention briefly.

From a theoretical point of view isofrequency pairing is of relevance
in view of perturbation techniques. For example, in order to use KAM
theory for perturbed integrable systems, certain non-degeneracy conditions
have to be satisfied. The simplest version of these non-degeneracy
conditions is obviously violated if there is a degeneracy in the frequencies
such that other, more complex or more restrictive, conditions have
to be tested. To mention another example, the feature of isofrequency
pairing and the occurrence of a singular curve can be used to compare
different approaches to the general relativistic two-body problem,
as it was already mentioned in \cite{Warburton13}. Methods such as
the effective one-body approach or the post-Newtonian approximation
can profit from the isofrequency pairing and its related characteristics.

Also from a theoretical point of view, it is an interesting question
to ask if there are spacetimes where three or more orbits with the
same frequencies exist. In all examples treated so far there are only
isofrequency \emph{pairs}. (Here we are refering to the situation
that all the parameters of the dynamical system have been fixed which,
for the cases treated in this paper, means fixing $\Lambda$ and $s$.)
As an attempt to find a candidate for isofrequency triples one could
start with a Bertrand spacetime and perturb it a little bit. Bertrand
spacetimes, which were introduced in \cite{Perlick92}, are spherically
symmetric and static spacetimes in which the ratio of the radial frequency
and the azimuthal frequency is a constant rational number $q$ for
all bound orbits, so they show the same total degeneracy of the frequencies
as the Kepler problem but now with $q\neq1$. We are planning to search
for isofrequency triples etc. in future work.



\section*{Acknowledgments}

We gratefully acknowledge support from the Deutsche Forschungsgemeinschaft
within the Research Training Group 1620 ``Models of Gravity'' and
from the ``Centre for Quantum Engineering and Space-Time Research
(QUEST)''. VP was financially supported by Deutsche Forschungsgemeinschaft,
Grant No. LA 905/14-1.

\section*{Appendix}

\subsection*{Approximation of hyperelliptic Integrals}

In this section we briefly describe how we use the approximation scheme
for hyperelliptic integrals developed by Sochnev \cite{Sochnev68}.
He based the method on the approximation of irrational functions by
rational ones. In particular, the irrational function $c=\sqrt[m]{c_{1}c_{2}...c_{m}}$
can be approximated by the sequences $\{a_{n}\}$ and $\{b_{n}\}$
which are defined iteratively by 
\begin{equation}
a_{1}=\frac{c_{1}+c_{2}+...+c_{m}}{m}\,,\qquad b_{1}=\frac{c_{1}c_{2}...c_{m}}{a_{1}^{m-1}}\,,
\end{equation}
and 
\begin{equation}
a_{n+1}=\frac{(m-1)a_{n}+b_{n}}{m}\,,\qquad b_{n+1}=\frac{a_{n}^{m-1}b_{n}}{a_{n+1}^{m-1}}
\end{equation}
for $n\ge1$. While $\{a_{n}\}$ approaches $c$ from above, $\{b_{n}\}$
comes from below, i.e. $a_{1}>a_{2}>...>a_{n}>c>b_{n}>...>b_{2}>b_{1}$,
and their common limit for $n\rightarrow\infty$ is $c$.

Let us consider a specific example related to our problem of hyperelliptic
integrals. The irrational function $\sqrt[m]{1+kx}$ with $|k|<1$
is finite for $-1<x<1$. (These boundaries become important when we
consider the boundaries of the integrals.) Therefore, we can approximate
this function by choosing $\sqrt[m]{1+kx}=\sqrt[m]{c_{1}c_{2}...c_{m}}$
with $c_{1}=1+kx$ and $c_{2}=c_{3}=...=c_{m}=1$ and evaluating the
sequences $\{a_{n}\}$ and $\{b_{n}\}$ within the defined range of
$x$. In this way we approximate our irrational function by rational
ones.

This method can be used for evaluating our hyperelliptic integrals
\eqref{eq:TrSochnev}, \eqref{eq:DeltaPhiSochnev} 
where $m=2$. 
To that end we have to approximate the function $\sqrt{V_{r}(x)}$
where $V_{r}(x)$ is given by (\ref{eq:Vrx}). If the absolute values
of the $k$ factors are far away from one, the procedure goes like
this: First, we extract the factor $(1-x^{2})$ out of the radicand,
arrange the $k$ factors in decreasing order in their absolute values
and group positive and negative $k$s. Then we form subgroups with
$m=2$ elements within each group where we supplement a factor of
one when there are less than $m=2$ elements in the subgroups, i.e.
\begin{gather}
\sqrt{V_{r}(x)}=\sqrt{1-x^{2}}\\
\times\sqrt{(1-k_{3}x)(1+k_{1}x)}\,\sqrt{(1+k_{2}x)\cdot1}\,\sqrt{(1+k_{4}x)\cdot1}\nonumber 
\end{gather}
%
Now we are able to compute the approximating sequences $\{a_{n}\}$
and $\{b_{n}\}$ for each subgroup up to arbitrarily high order in
$n$, always resulting in a rational function. Consequently, the integral
that has to be solved can be approximated by integrals of the form
\[
\int_{-1}^{1}R\left(x,\sqrt{1-x^{2}}\right)dx
\]
where $R\left(x,\sqrt{1-x^{2}}\right)$ denotes a rational function
of $x$ and $\sqrt{1-x^{2}}$. Using any of the three Euler substitutions
or elementary transformations which rearrange the form of the integral
into tabulated ones, it is possible to solve the integral in terms
of elementary functions.

This gives a good approximation scheme as long as the absolute values
of all $k$s are far away from one. However, we are interested in
frequencies close to the separatrices corresponding to absolute values
close to one for either $k_{3}$ or $k_{4}$. Luckily, only a few
modifications to the procedure are necessary to adapt it to this case.

To begin with the integrals in 
\eqref{eq:TrSochnev} and \eqref{eq:DeltaPhiSochnev} are divided
into two integrals, where one runs from $-1$ to $0$ and the other
from $0$ to $1$. Then, it is not the factor $(1-x^{2})$ that is
extracted. Consider 
\[
(1-x)(1+x)(1+k_{3}x)(1+k_{1}x)(1+k_{2}x)(1+k_{4}x)
\]
where the $k$ factors are already arranged in decreasing order in
their absolute values and $(1-x^{2})$ is rewritten as $(1+x)(1-x)$.
In the first integral (from $-1$ to $0$) the product $(1+x)(1+k_{3}x)$
is taken out where $k_{3}$ is the greatest of the positive coefficients
$k_{i}$ leading to 
\[
\int_{-1}^{0}\frac{dx}{\sqrt{(1+x)(1+k_{3}x)}\,\mathcal{A}(x,k_{1},k_{2},k_{4})}
\]
where {\footnotesize{}{}{}$\mathcal{A}(x,k_{1},k_{2},k_{3})=\sqrt{(1+k_{1}x)(1+k_{2}x)(1-x)(1+k_{4}x)}$}
has to be approximated by the same procedure as explained above. The
second integral (from $0$ to $1$) is rearranged in such a way that
it yields 
\[
\int_{0}^{1}\frac{dx}{\sqrt{(1-x)(1+k_{4}x)}\,\mathcal{B}(x,k_{1},k_{2},k_{3})}
\]
with {\small{}{}{}$\mathcal{B}(x,k_{1},k_{2},k_{3})=\sqrt{(1+x)(1+k_{3}x)(1+k_{1}x)(1+k_{2}x)}$}.
Here, the factor $(1+k_{4}x)$ is pulled out together with $(1-x)$
because $k_{4}$ is the greatest of the negative coefficients $k_{i}$.
Again, the remaining function is approximated resulting in a rational
function in $x$. Therefore we obtain for our integral an approximation
of the form 
\begin{align*}
\int_{-1}^{0}\frac{R_{1}(x)dx}{\sqrt{(1+x)(1+k_{3}x)}}+\int_{0}^{1}\frac{R_{2}(x)dx}{\sqrt{(1-x)(1+k_{4}x)}}
\end{align*}
the solution of which can be found in terms of elementary functions.
In order to simplify the calculations further we may apply partial
fraction decompositions to each of the rational functions providing
us with integrals of the form 
\begin{align*}
\int_{-1}^{0}\frac{dx}{(1+a(k_{1},k_{2},k_{4})x)\sqrt{(1+x)(1+k_{3}x)}}\,,\\
\int_{0}^{1}\frac{dx}{(1+b(k_{1},k_{2},k_{3})x)\sqrt{(1-x)(1+k_{4}x)}}\,.
\end{align*}
In the spinning case, we have to evaluate the integrals (\ref{eq:TrSpin})
and (\ref{eq:DeltaPhiSpin}), i.e., we have to approximate the function
$\sqrt{V_{r}^{s}(x)}$ with $V_{r}^{s}(x)$ given by (\ref{eq:Vrsx}).
It is convenient to treat the two non-real zeros and the real zeros
separately. Since 
\[
\sqrt{\left(x-\left(a_{I}+ib_{I}\right)\right)\left(x-\left(a_{I}-ib_{I}\right)\right)}
\]
is a real-valued irrational function it is possible to approximate
it by Sochnev's method leading to 
\begin{align*}
\sqrt{\left(x-\left(a_{I}+ib_{I}\right)\right)\left(x-\left(a_{I}-ib_{I}\right)\right)} & \rightarrow\\
\left(\frac{r_{a}-r_{p}}{2}-a_{I}\right)\left(1+\frac{r_{a}-r_{p}}{r_{a}+r_{p}-2a_{I}}x\right)
\end{align*}
in first approximation, which proves to be of sufficient accuracy
for our purposes. Therewith, the remaining terms {\scriptsize{}{}{}
\[
\sqrt{(1-x^{2})(1+k_{1}x)(1+k_{2}x)(1+k_{3}x)(1+k_{4}x)(1+k_{5}x)(1+k_{6}x)}
\]
}can be approximated analogously to the non-spinning case. Here, close
to the separatrices either $k_{2}$ or $-k_{3}$ approaches $1$,
and the approximation only contains elementary integrals which, after
partial fraction decomposition, reduce to integrals of the form 
\begin{align}
\int_{-1}^{0}\frac{dx}{(1+a_{s}(k_{1},k_{2},k_{4})x)\sqrt{(1+x)(1+k_{2}x)}}\,,\nonumber \\
\int_{0}^{1}\frac{dx}{(1+b_{s}(k_{1},k_{2},k_{3})x)\sqrt{(1-x)(1+k_{3}x)}}\,.\label{eq:SochnevIntegralsSpin}
\end{align}

\end{document}